\documentclass{optica-article}

\journal{opticajournal} 

\articletype{Research Article}

\usepackage{lineno}
\usepackage{booktabs}
\usepackage{subfigure}

\begin{document}

\title{Broadband switchable infrared absorbers using phase-change materials}

\author{Md Tanvir Emrose,\authormark{1,2} Emily L. Payne,\authormark{3} 
Chenglong You,\authormark{4}
and Georgios Veronis\authormark{1,2,*}}

\address{\authormark{1}School of Electrical Engineering and Computer Science, Louisiana State University, Baton Rouge, Louisiana 70803, USA\\
\authormark{2}Center for Computation and Technology, Louisiana State University, Baton Rouge, Louisiana 70803, USA\\
\authormark{3}Department of Physics, Brigham Young University - Idaho, Rexburg, Idaho, 83460, USA\\
\authormark{4}Department of Physics and Astronomy, Louisiana State University, Baton Rouge, Louisiana 70803, USA}

\email{\authormark{*}gveronis@lsu.edu} 


\begin{abstract*} 
We introduce multilayer structures with the phase-change material Ge$_2$Sb$_2$Te$_5$ (GST) for use as broadband switchable absorbers in the infrared wavelength range.
We optimize both the material composition and the layer thicknesses of the multilayer structures, in order to maximize the difference between the absorption for GST in its crystalline phase and the absorption for GST in its amorphous phase in the wavelength range of interest. 
We show that in the optimized structures near perfect absorption can be switched to very low absorption in a broad wavelength range by switching GST from its crystalline to its amorphous phase.
Our optimized lithography-free structures have better performance than harder-to-fabricate three-dimensional structures.
Our results could pave the way to a new class of broadband switchable absorbers and thermal sources in the infrared wavelength range.
\end{abstract*}

\section{Introduction}

Controlling the transmission, reflection, and absorption properties of optical nanostructures has been of great interest in recent years. In particular, controlling absorption is important for applications such as thin-film thermal emitters, color displays, photovoltaics, 
and smart windows \cite{so2021_9,jafari_2019_10,wang2012_7-3,mkhitaryan2017_22}. Infrared light absorbers play a crucial role in biochemical sensing, thermal imaging, solar energy harvesting, and thermal management \cite{liu2010_7-1, tittl2015_6-8, polman2016_6-7, hossain2015_6-6, prakashS2022_7, badloe2020-8}. 
In addition, achieving near complete light absorption in the mid-infrared spectral range is especially important for industrial applications because this range includes wavelengths where objects with 300-700 °C temperatures exhibit peak thermal emission \cite{heblar2021_6}.

Materials with tunable absorption coefficient such as phase-change materials (PCMs), can be used to design tunable and reconfigurable absorbers. PCMs exhibit several desirable characteristics such as high-speed phase switching with minimal power consumption, long-term stability, and a large contrast between the optical properties of their two distinct phases \cite{abdollahramezani2020_review-2}. 
PCMs can be switched between their two structural phases via thermal annealing, electrical pulses, or optical pulses \cite{abdollahramezani2021_3}. The transition between the two phases can be at subpicosecond timescales \cite{Rude:16,Loke:12}.  
In addition, many PCMs exhibit low losses in the infrared region, rendering them suitable for various nanophotonics applications, including photonic memory \cite{mandal_2021_review-1}, optical limiters \cite{sarangan2018_19}, cloaking devices \cite{huang2018_63}, optical switches \cite{huang2017_60}, optical modulators \cite{chamoli2020_4}, tunable metasurfaces \cite{abdollahramezani2021_3, wang2021_15, pogrebnyakov2018_16}, and switchable absorbers \cite{heblar2021_6, prakashS2022_7, badloe2020-8, tittl2015_6-8}. 

In photonic applications, two commonly used PCMs are germanium-antimony-tellurium (GST) and vanadium dioxide (VO$_2$). VO$_2$ undergoes a reversible phase change from an insulating to a metallic phase \cite{badloe2020-8,raoux2009_19-7}. 
VO$_2$ is volatile, requiring a continuous power supply to maintain its metallic phase. GST, which is a chalcogenide glass, undergoes an amorphous to crystalline phase transition with large changes in its dielectric constant at fast time scales \cite{siegel2004_19-8}. 
In addition, the phase change in GST is nonvolatile and, therefore, power is consumed only during the phase transition process \cite{huang2022_80}.
GST can be deposited on virtually any substrate or film at low temperatures, allowing its integration into complex multilayer structures \cite{sarangan2018_19}. 
These features make GST an ideal candidate for reconfigurable infrared absorbers.
Crystallization in Ge$_2$Sb$_2$Te$_5$ and its stoichiometric neighbors Ge$_2$Sb$_1$Te$_4$, Ge$_2$Sb$_2$Te$_4$, Ge$_3$Sb$_4$Te$_8$, and GeTe, leads to an approximately twofold increase in the refractive index within the infrared spectral range \cite{nature_date}. 
This large contrast is associated with resonant bonding effects \cite{nature_date}.
Among GST materials, Ge$_2$Sb$_2$Te$_5$ is the preferred choice for photonic devices due to its exceptional characteristics, including ultrafast switching and high refractive index contrast between its two phases \cite{abdollahramezani2020_review-2}.

In this paper, we introduce multilayer structures with the phase-change material GST (Ge$_2$Sb$_2$Te$_5$) above a semi-infinite metal substrate for use as broadband switchable absorbers in the infrared wavelength range. We use a memetic optimization algorithm to optimize both the material composition and the layer thicknesses of the multilayer structures, in order to maximize the difference between the absorption for GST in its crystalline phase and the absorption for GST in its amorphous phase in the wavelength range of interest. 
We first employ the memetic optimization algorithm to design a switchable absorber operating in the 1-1.6 $\mu$m wavelength range. Our optimized five-layer structure consists of a thin GST film on top of a MgF$_2$ dielectric spacer backed with a silver mirror. The structure also includes three additional dielectric layers on top of the GST layer.
We find that the average absorption difference between the two GST phases in the 1-1.6 $\mu$m wavelength range is $\sim$67.8\%.
In addition, although the multilayer structure is optimized for normally incident light, it achieves large absorption difference between the two GST phases in a broad angular range.
Our optimized lithography-free five-layer structure has better performance than three-dimensional structures reported in the literature which require nanoimprint or electron beam lithography.
We also employ the memetic optimization algorithm to design a switchable absorber operating in the 2.1-4.1 $\mu$m wavelength range. Our optimized five-layer structure includes a thin and a thick GST layer, separated by a SiC layer. 
In addition, a MgF$_2$ dielectric layer separates the bottom GST layer from the silver substrate, while an HfO$_2$ layer separates the top GST layer from air.
Amorphous GST is almost lossless for wavelengths longer than 2 $\mu$m, while crystalline GST is lossy in the same wavelength range leading to superior performance of the 2.1-4.1 $\mu$m switchable absorber compared to the 1-1.6 $\mu$m switchable absorber.
We find that the average absorption difference between the two GST phases in the 2.1-4.1 $\mu$m wavelength range is $\sim$97.2\%. 
Thus, our optimized structure approaches the ideal structure, since it is almost a perfect absorber for GST in its crystalline phase and almost a perfect reflector for GST in its amorphous phase in the 2.1-4.1 $\mu$m wavelength range of interest.

The remainder of the paper is organized as follows. In Section 2, we introduce the merit function
involving the absorption difference between the two GST phases, and describe the memetic algorithm that we use for the optimization of the broadband switchable absorbers.
In Subsections 3.1 and 3.2, we employ the memetic optimization algorithm to design switchable absorbers operating in the 1-1.6 $\mu$m and 2.1-4.1 $\mu$m wavelength ranges, respectively. Finally, our conclusions are summarized in Section 4.

\section{Theory}
\label{Theory}

We consider aperiodic multilayer structures with the phase-change material GST above a semi-infinite metal substrate.  
Multilayer structures can provide absorption spectra similar to that of more complex and harder-to-fabricate three-dimensional structures \cite{granier2014optimized}.
Light is incident from air at an angle $\theta$. We use experimental data for the wavelength-dependent refractive indices of all materials considered \cite{palik1998_5-29, kischkat2012_5-30, nature_date}. Since light cannot be transmitted through the metal substrate, the absorption can be calculated using
\begin{equation}
    A_{\rm TE/TM}(\lambda, \theta)=1-R_{\rm TE/TM}(\lambda, \theta),
\end{equation}
where $A_{\rm TE}$ ($A_{\rm TM}$) is the absorption for TE (TM) polarization, $R_{\rm TE}$ ($R_{\rm TM}$) is the TE (TM) reflection, and $\lambda$ is the wavelength.
The reflection is calculated using the impedance method \cite{haus1984waves}.

In a structure with $l$ layers with thicknesses $\mathbf{d} = [d_1\, d_2 \, ... \, d_l]$ the material composition of the layers is represented by the vector $\mathbf{m} = [m_1 \, m_2 \, ...\, m_l]$, where $m_i \,(i\,=\,1,\,2,\,...\,,l)$ are integers between 1 and $M$, and $M$ is the number of materials considered.
Here, the material of each layer is chosen among the following nine materials ($M=9$): aluminum oxide (Al$_2$O$_3$), hafnium oxide (HfO$_2$), magnesium fluoride (MgF$_2$), silicon carbide (SiC), silicon nitride (Si$_3$N$_4$), silicon dioxide (SiO$_2$), tantalum pentoxide (Ta$_2$O$_5$), titanium dioxide (TiO$_2$), which are commonly used dielectrics, and GST, which is a phase change material. 

In general, we wish to achieve a specific difference between the absorption spectra for GST in its crystalline and amorphous phases.
We therefore first specify the target difference $D_{\mathrm{target}}(\lambda, \theta)$ between the absorption for GST in its crystalline phase and the absorption for GST in its amorphous phase.
We then minimize the merit function $F(\mathbf{m},\mathbf{d})$ involving the absorption difference between the two GST phases and the target absorption difference $D_{\mathrm{target}}(\lambda, \theta)$: 
\begin{equation}
\label{eq2}
    F(\mathbf{m},\mathbf{d}) =  \sum_{\lambda, \theta}^{} W(\lambda,\theta)[A_{\mathrm{cGST}}(\lambda, \theta; \mathbf{m},\mathbf{d}) - A_{\mathrm{aGST}}(\lambda, \theta; \mathbf{m},\mathbf{d}) - D_{\mathrm{target}}(\lambda, \theta)]^2,
\end{equation}
where $W(\lambda,\theta)$ is the weight associated with each wavelength-incident angle pair, and $A_{\mathrm{cGST}}(\lambda, \theta; \mathbf{m},\mathbf{d})$, $A_{\mathrm{aGST}}(\lambda, \theta; \mathbf{m},\mathbf{d})$ are the absorption for GST in its crystalline and amorphous phases, respectively.

In this work, we wish to maximize the difference between the absorption for GST in its crystalline phase and the absorption for GST in its amorphous phase in the wavelength range of interest. In other words, the ideal structure will have complete absorption for GST in its crystalline phase ($A_{\mathrm{cGST}}=1$), and complete reflection for GST in its amorphous phase ($A_{\mathrm{aGST}}=0$). We therefore choose the target difference to be $D_{\mathrm{target}}(\lambda, \theta)=1$.

We use a memetic optimization algorithm coupled with the impedance method to optimize both the material composition $\mathbf{m}$ and the layer thicknesses $\mathbf{d}$ of the aperiodic multilayer structures, in order to minimize the merit function $F(\mathbf{m},\mathbf{d})$ [Eq. (2)].
It has been shown that simultaneously optimizing the material composition as well as the layer thicknesses leads to structures which achieve better performance with smaller number of layers \cite{fan2017,you2020_72}.
The memetic algorithm combines an evolutionary global optimization method and a local optimization method.
The operations of the evolutionary algorithm 
are patterned after the natural selection process.
We modified the implementation of the memetic algorithm developed by Shi \textit{et al.} \cite{fan2017} in order to add phase change materials with two different phases to the list of potential materials for each layer, and to use a merit function which depends on the optical properties of both phases of such materials [Eq. (2)].

The memetic optimization algorithm implementation that we use involves several steps \cite{fan2017}. First, we randomly generate an initial population of $N$ $l$-layer structures. We then perform a crossover operation by pairing individuals as parents, and combining multilayer sections of the parents to create new children. We next perform a mutation operation by randomly varying the material and thickness of a layer in a randomly chosen individual. We then reselect the next generation of the population based on the merit function of individual structures [Eq. (2)], with a certain percentage of individuals chosen from the top and bottom percentiles of the population to ensure diversity. Elite individuals periodically undergo refinement through local optimization of their layer thicknesses using the quasi-Newton method. We repeat these steps until the population becomes sufficiently uniform and convergence is reached.
In order to achieve the minimization of the merit function [Eq. (2)], we also made an additional modification to the memetic algorithm implementation of Shi \textit{et al.}: Once convergence is reached, we retain the best individual of the population, and randomly generate $N-1$ new structures to form a new population and restart the evolution process. We repeat this process of retaining the best individual and randomly generating the rest of the population $P$ times.
We heuristically choose $N$ and $P$ to accelerate the convergence of the algorithm with typical values $N=2000$, $P=200$. We choose other parameters of the memetic optimization algorithm as in Shi \textit{et al.} \cite{fan2017}.

\section{Results}
\label{Results}

\begin{figure}[htb!]
\centering\includegraphics[width=11cm]{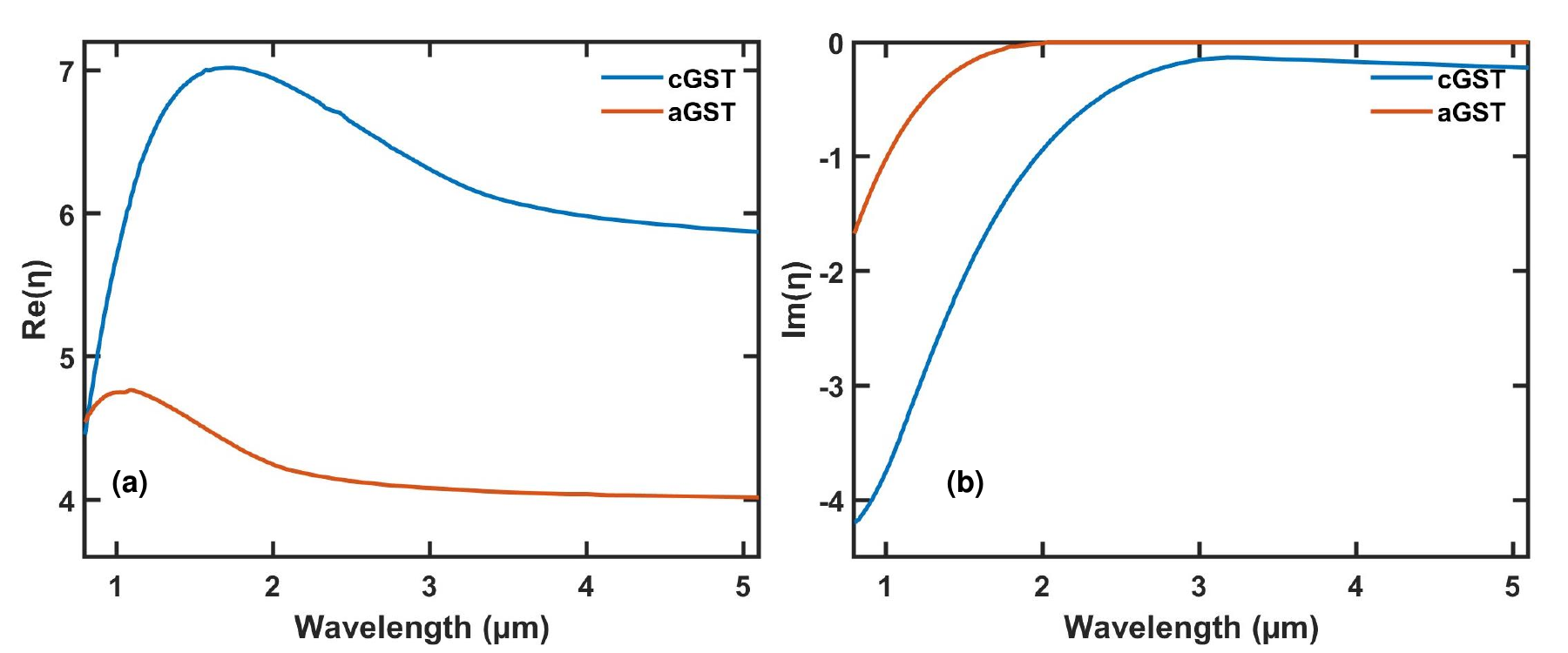} 
\caption{(a) Real and (b) imaginary part of the refractive index of GST in its amorphous (aGST) and crystalline (cGST) phases, as a function of wavelength.}
\label{fig:one}
\end{figure}

The optical properties of GST change significantly when transitioning between its amorphous and crystalline phases. This phase transition can be induced rapidly and reversibly through external electrical pulses, optical pulses, or thermal annealing, with power consumption occurring only during the transition process \cite{nature_date}. Figs. \ref{fig:one}(a) and \ref{fig:one}(b) show the real and imaginary part of the refractive index of GST in its amorphous (aGST) and crystalline (cGST) phases, as a function of wavelength \cite{nature_date}. We observe that the crystalline phase exhibits higher real part of the refractive index compared to the amorphous phase.  In addition, GST in its amorphous phase is almost lossless for wavelengths longer than 2 $\mu$m, since
the imaginary part of its refractive index is close to zero [Fig. \ref{fig:one}(b)]. 

In this work, we introduce multilayer structures with the phase-change material GST for use as broadband switchable absorbers in the infrared wavelength range. More specifically, we design switchable absorbers operating in two broad wavelength ranges in the infrared: 1-1.6 $\mu$m and 2.1-4.1 $\mu$m. We optimize the structures for normally incident light ($\theta=0^\circ$), and use equal weights $W(\lambda,\theta)$ for all wavelengths to calculate the merit function [Eq. (2)].

\subsection{Switchable absorber in the 1-1.6 $\mu$m wavelength range}
\label{1to1.6}

We first employ the memetic optimization algorithm described in Section 2 to design a switchable absorber operating in the 1-1.6 $\mu$m wavelength range. 
We optimize both the material composition and the layer thicknesses of a five-layer structure in order to maximize the difference between the absorption for GST in its crystalline phase and the absorption for GST in its amorphous phase in the 1-1.6 $\mu$m wavelength range for normally incident light by minimizing the merit function [Eq. (2)].
We consider  tungsten (W), silver (Ag), and aluminum (Al), which are commonly used metals in photonic devices, for the material of the substrate. We find that silver substrate leads to switchable absorbers with better performance in terms of the average absorption difference in the wavelength range of interest than tungsten or aluminum. We therefore choose silver as the substrate.
As mentioned above, the material of each of the five layers is chosen among Al$_2$O$_3$, HfO$_2$, MgF$_2$, SiC, Si$_3$N$_4$, SiO$_2$, Ta$_2$O$_5$, TiO$_2$, and GST, while the superstrate is air. 

\begin{figure}[htb!]
\centering
\centering\includegraphics[width=11cm]{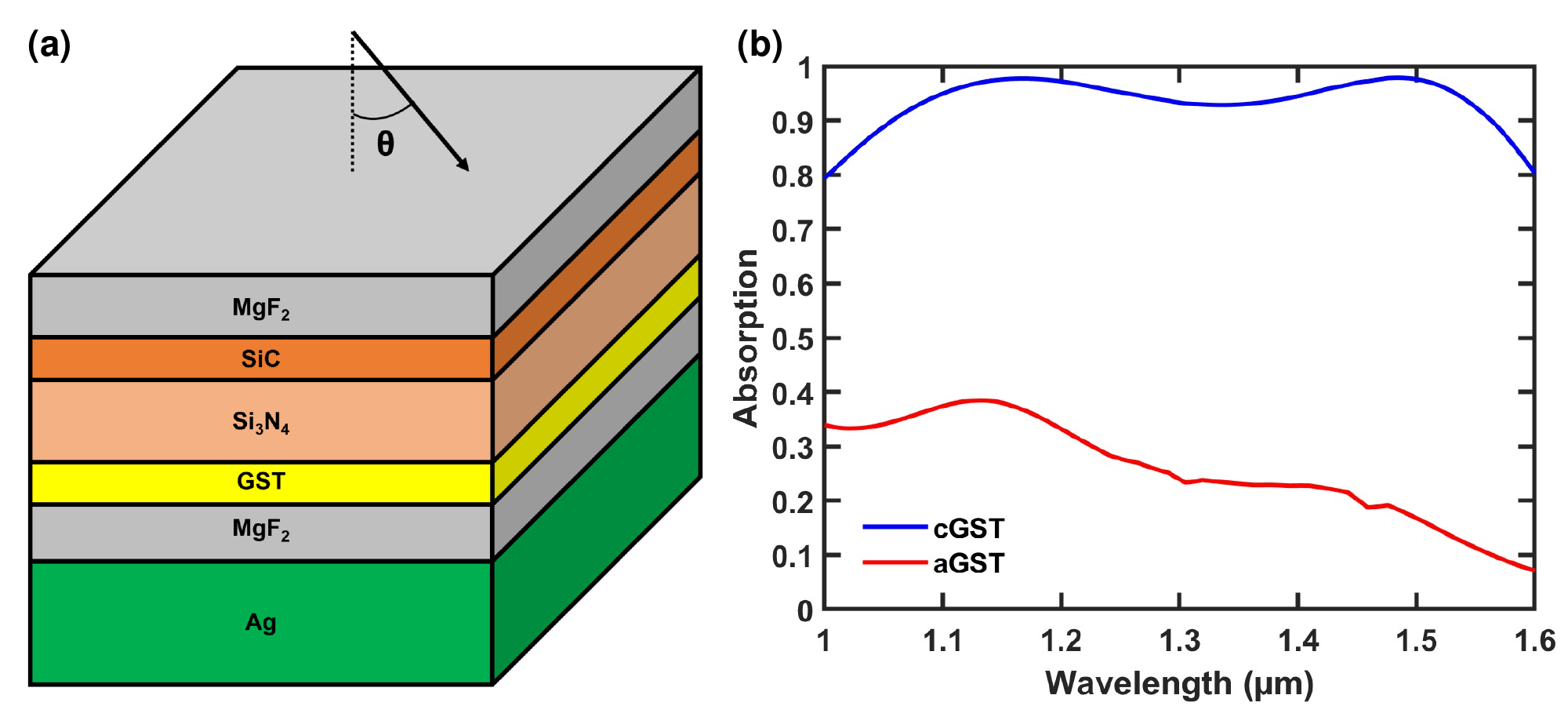} 
\caption{(a) Schematic showing the optimized material composition for a five-layer structure above a silver substrate. 
Here, we maximize the difference between the absorption for GST in its crystalline phase and the absorption for GST in its amorphous phase in the 1-1.6 $\mu$m wavelength range.
(b) Absorption as a function of wavelength for the optimized five-layer structure with GST in its crystalline and amorphous phases. Results are shown for normally incident light. The material and thickness of each layer in the optimized structure are given in Table \ref{table:one}.}
\label{fig:two}
\end{figure}

In Fig. 2(a), we show the optimized material composition for the five-layer structure above the silver substrate obtained from the memetic optimization algorithm for a switchable absorber in the 1-1.6 $\mu$m wavelength range. 
Table 1 shows the material and thickness of each layer in the optimized five-layer structure of Fig. \ref{fig:two}(a).
The optimized structure consists of a thin GST film on top of a MgF$_2$ dielectric spacer backed with a Ag mirror (Table \ref{table:one}). The structure also includes three additional dielectric layers (MgF$_2$-SiC-Si$_3$N$_4$) on top of the GST layer.

In Fig. \ref{fig:two}(b), we show the absorption for normally incident light ($\theta=0^\circ$) as a function of wavelength for the optimized five-layer structure of Table 1 with GST in its crystalline and amorphous phases. We find that the average absorption in the 1-1.6 $\mu$m wavelength range for GST in its crystalline phase is $\sim$93.6\%, while the minimum absorption in this range is $\sim$79.5\%. 
On the other hand, for GST in its amorphous phase the average absorption in the 1-1.6 $\mu$m wavelength range is reduced to $\sim$25.8\%, while the maximum absorption in this range is $\sim$38.5\%.
Our optimized lithography-free five-layer structure has therefore an average absorption difference between the two GST phases of $\sim$67.8\% in the 1-1.6 $\mu$m wavelength range which is better than the one of three-dimensional structures reported in the literature which require nanoimprint or electron beam lithography \cite{badloe2020-8}.

\begin{table}[htb!]
    \centering
    \caption{Material and Thickness of Each Layer in the Optimized Structure of Fig. \ref{fig:two}(a)}
    \begin{tabular}{c|c|c|}
        \toprule
        Layer   & Material & Thickness ($\mu$m)\\
        \midrule
                & Air       & Superstrate\\
        1       & MgF$_2$   & 0.670\\
        2       & SiC       & 0.252\\
        3       & Si$_3$N$_4$       & 0.073\\
        4       & GST       & 0.008\\
        5       & MgF$_2$   & 0.078\\
                & Ag        & Substrate\\
        \bottomrule
    \end{tabular}
    \label{table:one}
\end{table}

\begin{figure}[htb!]
\centering
\centering\includegraphics[width=12cm]{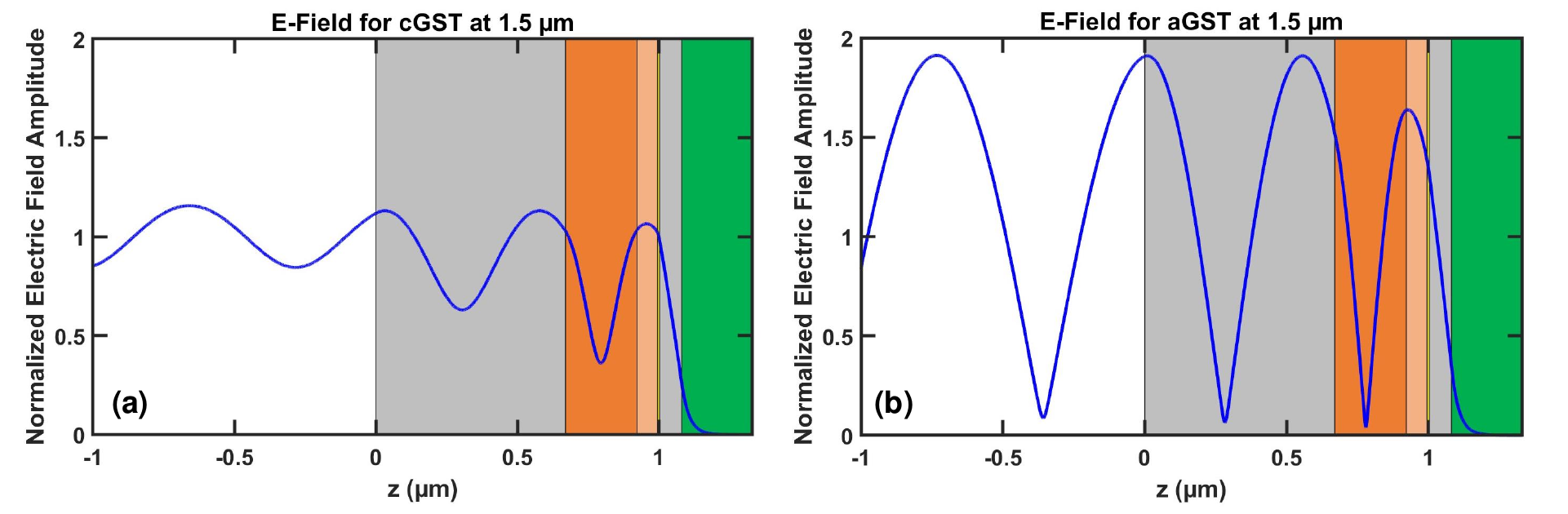} 
\caption{
Profile of the electric field amplitude, normalized with respect to the field amplitude of the normally incident plane wave, with $\lambda$ = 1.5 $\mu$m for the optimized five-layer structure of Table 1. Results are shown for GST in its (a) crystalline and (b) amorphous phases. For GST in its crystalline phase, the ratio of the power absorbed in each layer to the total absorbed power, beginning with the layer adjacent to air, is \{0, 0, 0, 0.97, 0, 0.03\}. In other words, $\sim$97\% of the total power absorbed is absorbed in the GST layer, while $\sim$3\% is absorbed in the silver substrate. For GST in its amorphous phase, the ratio of the power absorbed in each layer to the total absorbed power is \{0, 0, 0, 0.67, 0, 0.33\}.
}
\label{fig:three}
\end{figure}

In Figs. \ref{fig:three}(a) and \ref{fig:three}(b), we show the profile of the electric field amplitude, normalized with respect to the field amplitude of the normally incident plane wave ($\lambda$ = 1.5 $\mu$m), for the optimized five-layer structure of Table 1, with GST in its crystalline and amorphous phases, respectively.
We observe that, with GST in its crystalline phase, there is almost perfect impedance matching between the structure and air, resulting in a very low reflection of $\sim$2\% [Fig. \ref{fig:three}(a)]. On the other hand, when GST is switched to its amorphous phase, the profile of the electric field amplitude changes drastically, and the reflection increases to $\sim$83\% [Fig. \ref{fig:three}(b)].
In other words, despite the small thickness of the GST layer in the optimized structure (Table \ref{table:one}), switching between the crystalline and amorphous phases of GST leads to drastic changes in the electric field profile.

\begin{figure}[htb!]
\centering
\centering\includegraphics[width=11cm]{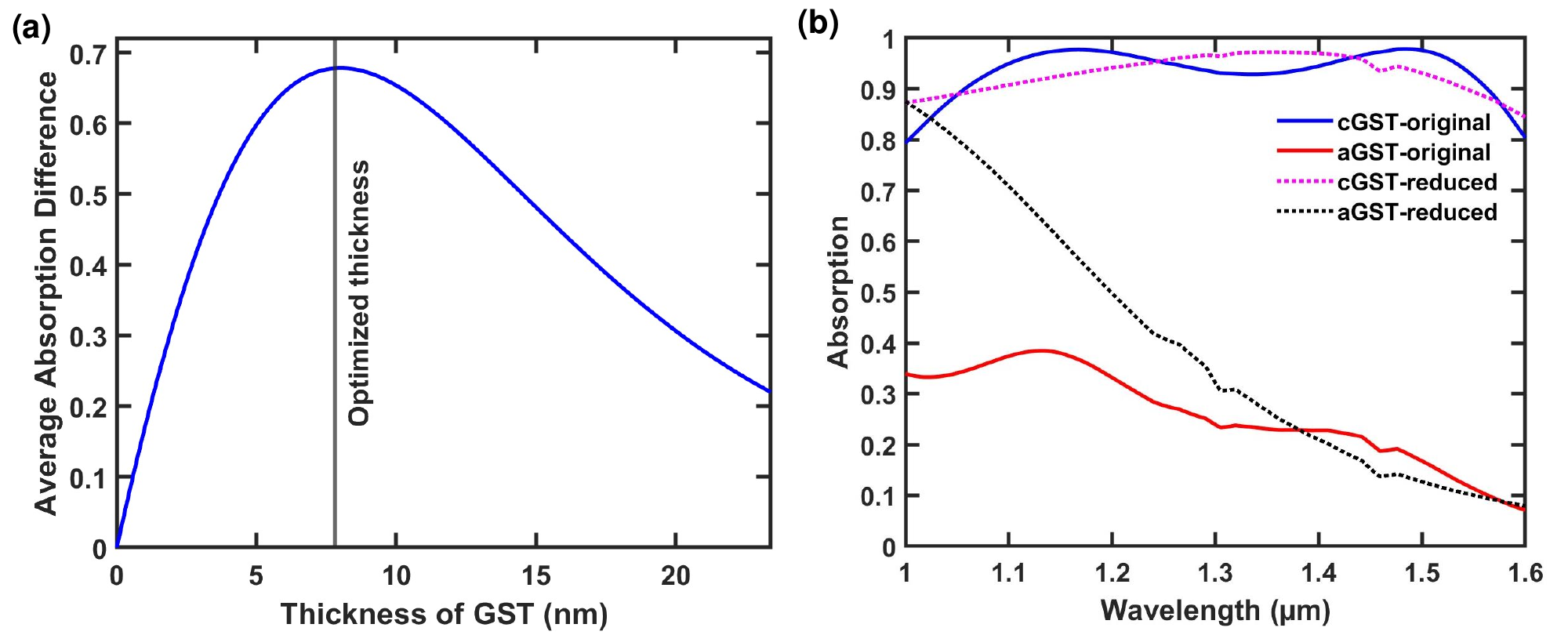} 
\caption{
(a) Average difference in the 1-1.6 $\mu$m wavelength range between the absorption for GST in its crystalline phase and the absorption for GST in its amorphous phase as a function of the GST layer thickness. All other parameters are as in Table \ref{table:one}. Results are shown for normally incident light.
(b) Absorption as a function of wavelength for the optimized five-layer structure of Table \ref{table:one} with GST in its crystalline (cGST-original) and amorphous (aGST-original) phases. We also show the absorption for a reduced two-layer GST-MgF$_2$ structure with GST in its crystalline (cGST-reduced) and amorphous (aGST-reduced) phases. 
The reduced structure is obtained by removing the first three layers adjacent to air from the optimized structure of Table \ref{table:one}, while the thicknesses of the remaining two layers are as in Table \ref{table:one}. 
Results are shown for normally incident light.
}
\label{fig:four}
\end{figure}

Figure~\ref{fig:four}(a) shows the average difference in the 1-1.6 $\mu$m wavelength range between the absorption for GST in its crystalline phase and the absorption for GST in its amorphous phase as a function of the GST layer thickness. We observe that the average absorption difference is highly dependent on the thickness of the thin GST layer. The optimized GST layer thickness is 8 nm (Table \ref{table:one}).

The absorption in a thin lossy film can drastically increase by putting the film on top of a dielectric spacer backed with a mirror \cite{piper2014, tischler2006}. Our optimized structure indeed consists of a thin lossy GST film on top of a MgF$_2$ dielectric spacer backed with a Ag mirror (Table \ref{table:one}). However, our optimized structure also includes a MgF$_2$-SiC-Si$_3$N$_4$ three-layer structure on top of the GST layer (Table \ref{table:one}). To elucidate the role of these three layers on top of the GST layer, in Fig. \ref{fig:four}(b) we show the absorption as a function of wavelength with GST in its crystalline (cGST-reduced) and amorphous (aGST-reduced) phases for a reduced two-layer GST-MgF$_2$ structure. The reduced structure is obtained by removing the first three layers adjacent to air from the optimized five-layer structure of Table \ref{table:one}. For comparison, we also show the absorption of the optimized structure with GST in its crystalline (cGST-original) and amorphous (aGST-original) phases. For GST in its crystalline phase (cGST), the absorption of the reduced two-layer structure is very similar to the absorption of the optimized five-layer structure [Fig. \ref{fig:four}(b)]. However, for GST in its amorphous phase (aGST) the absorption of the reduced two-layer structure increases as the wavelength decreases, and the absorption difference between the two phases therefore decreases [Fig. \ref{fig:four}(b)]. Thus, we conclude that the three dielectric layers on top of the GST layer enable our optimized five-layer structure of Table \ref{table:one} to achieve large absorption difference between the two GST phases in a broader wavelength range.
In addition, we found that these three dielectric layers also lead to large absorption difference between the two GST phases in a broader angular range.

\begin{figure}[htb!]
\centering
\centering\includegraphics[width=11cm]{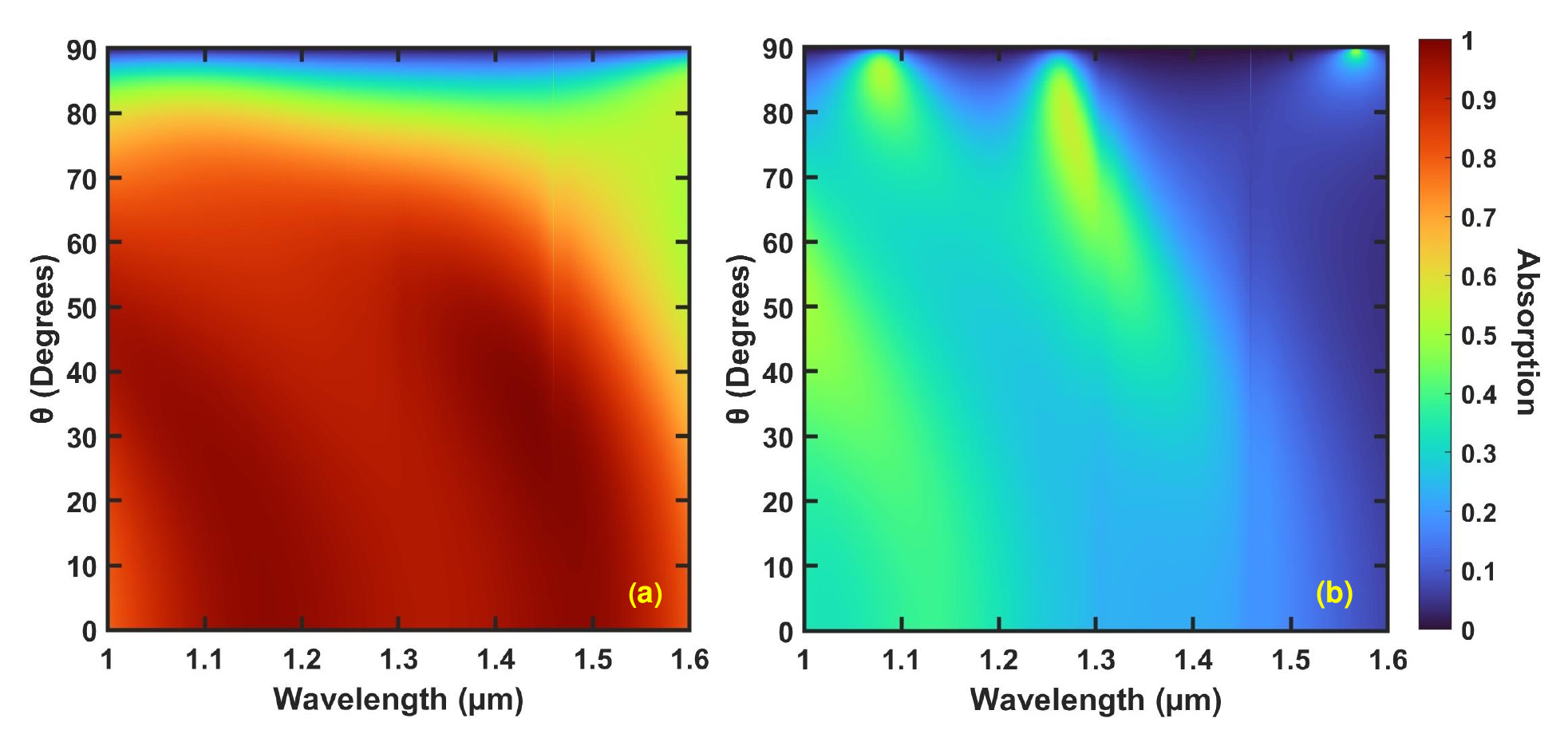} 
\caption{
Absorption as a function of wavelength and angle for the optimized five-layer structure of Table \ref{table:one} with GST in its (a) crystalline and (b) amorphous phases.
}
\label{fig:five}
\end{figure}

As mentioned above, we optimize the multilayer structure to maximize the difference between the absorption for GST in its crystalline phase and the absorption for GST in its amorphous phase in the 1-1.6 $\mu$m wavelength range for normally incident light. Here, we also investigate the dependence of the absorption difference on the angle of incidence. In Figs. \ref{fig:five}(a) and \ref{fig:five}(b) we show the absorption as a function of wavelength and angle of incidence for the optimized five-layer structure of Table 1 with GST in its crystalline and amorphous phases, respectively. We calculate the absorption for unpolarized light by averaging the absorption over the TE and TM polarizations \cite{fan2017, you2020_72}. Since the optimized structure was designed to be broadband, we expect that the absorption spectra will not have a strong angular dependence, since the structure is inherently nonresonant \cite{fan2017}. We indeed observe that, although the multilayer structure was optimized for normally incident light, the difference between the absorption for GST in its crystalline phase [Fig. \ref{fig:five}(a)] and the absorption for GST in its amorphous phase [Fig. \ref{fig:five}(b)] is high in a broad angular range within the 1-1.6 $\mu$m wavelength range of interest. In other words, the proposed structures achieve large absorption difference between the two GST phases both in a broad wavelength and in a broad angular range.

\subsection{Switchable absorber in the 2.1-4.1 $\mu$m wavelength range}
\label{2.1to4.1}

We also employ the memetic optimization algorithm described in Section 2 to design a switchable absorber operating in the 2.1-4.1 $\mu$m wavelength range. 
As in the previous subsection, we optimize both the material composition and the layer thicknesses of a five-layer structure in order to maximize the absorption difference between the two GST phases in the wavelength range of interest for normally incident light.

In Fig. 6(a), we show the optimized material composition for the five-layer structure above the silver substrate obtained from the memetic optimization algorithm for a switchable absorber in the 2.1-4.1 $\mu$m wavelength range. 
Table 2 shows the material and thickness of each layer in the optimized five-layer structure of Fig. 6(a).
The optimized structure includes a thin and a thick GST layer with thicknesses of 0.035 $\mu$m and 8.104 $\mu$m, respectively, separated by a SiC layer. In addition, a MgF$_2$ dielectric layer separates the bottom GST layer from the silver substrate, while an HfO$_2$ layer separates the top GST layer from air.

\begin{figure}[htb!]
\centering
\centering\includegraphics[width=11cm]{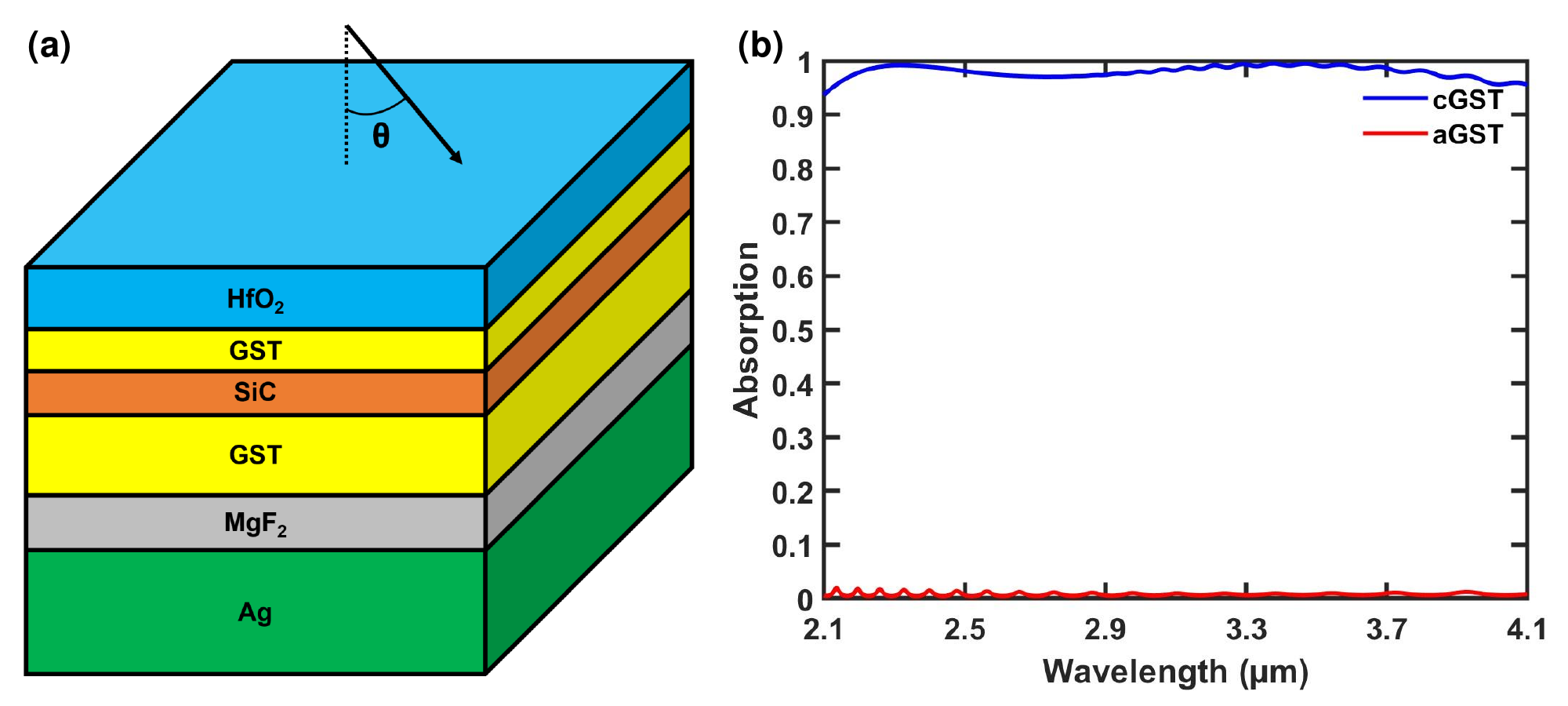} 
\caption{
(a) Schematic showing the optimized material composition for a five-layer structure above a silver substrate.
Here, we maximize the difference between the absorption for GST in its crystalline phase and the absorption for GST in its amorphous phase in the 2.1-4.1 $\mu$m wavelength range.
(b) Absorption as a function of wavelength for the optimized five-layer structure with GST in its crystalline and amorphous phases. Results are shown for normally incident light. The material and thickness of each layer in the optimized structure are given in Table \ref{table:two}.
}
\label{fig:six}
\end{figure}

In Fig.~\ref{fig:six}(b), we show the absorption for normally incident light ($\theta=0^\circ$) as a function of wavelength for the optimized five-layer structure of Table 2 with GST in its crystalline and amorphous phases. 
We find that the average absorption in the 2.1-4.1 $\mu$m wavelength range for GST in its crystalline phase is $\sim$98.0\%, while the minimum absorption in this range is $\sim$93.6\%. 
On the other hand, for GST in its amorphous phase the average absorption in the 2.1-4.1 $\mu$m wavelength range is reduced to $\sim$0.8\%, while the maximum absorption in this range is $\sim$1.8\%. In other words, the optimized structure of Table \ref{table:two} approaches the ideal structure, since it is almost a perfect absorber for GST in its crystalline phase and almost a perfect reflector for GST in its amorphous phase in the 2.1-4.1 $\mu$m wavelength range of interest.

The optimized five-layer switchable absorber operating in the 2.1-4.1 $\mu$m wavelength range (Table 2) has a $\sim$97.2\% average absorption difference between the two GST phases, which is significantly larger than the $\sim$67.8\% average absorption difference of the optimized switchable absorber operating in the 1-1.6 $\mu$m wavelength range (Table 1). The superior performance of the 2.1-4.1 $\mu$m switchable absorber is mostly due to its significantly lower $\sim$0.8\% average absorption for amorphous GST [Fig. 6(b)], compared to the $\sim$25.8\% average absorption of the 1-1.6 $\mu$m switchable absorber [Fig. 2(b)].
We found that the superior performance of the 2.1-4.1 $\mu$m switchable absorber for amorphous GST is directly related to the optical properties of GST (Fig. 1). More specifically amorphous GST is almost lossless for wavelengths longer than 2 $\mu$m, since the imaginary part of its refractive index is close to zero [Fig. 1(b)], while crystalline GST is lossy in the same wavelength range. This enables us to design a close to ideal broadband switchable absorber for wavelengths longer than 2 $\mu$m, with almost perfect absorption for the lossy crystalline GST, and almost perfect reflection for the lossless amorphous GST. On the other hand, for wavelengths shorter than 2 $\mu$m both GST phases are lossy, and it is therefore hard to achieve close to ideal performance.

\begin{table}[htb!]
    \centering
    \caption{Material and Thickness of Each Layer in the Optimized Structure of Fig. \ref{fig:six}(a)}
    \begin{tabular}{c|c|c|}
        \toprule
        Layer   & Material & Thickness ($\mu$m)\\
        \midrule
                & Air       & Superstrate\\
        1       & HfO$_2$   & 0.353\\
        2       & GST       & 0.035\\
        3       & SiC       & 0.070\\
        4       & GST       & 8.104\\
        5       & MgF$_2$   & 0.491\\
                & Ag        & Substrate\\
        \bottomrule
    \end{tabular}
    \label{table:two}
\end{table}

\begin{figure}[htb!]
\centering
\centering\includegraphics[width=12cm]{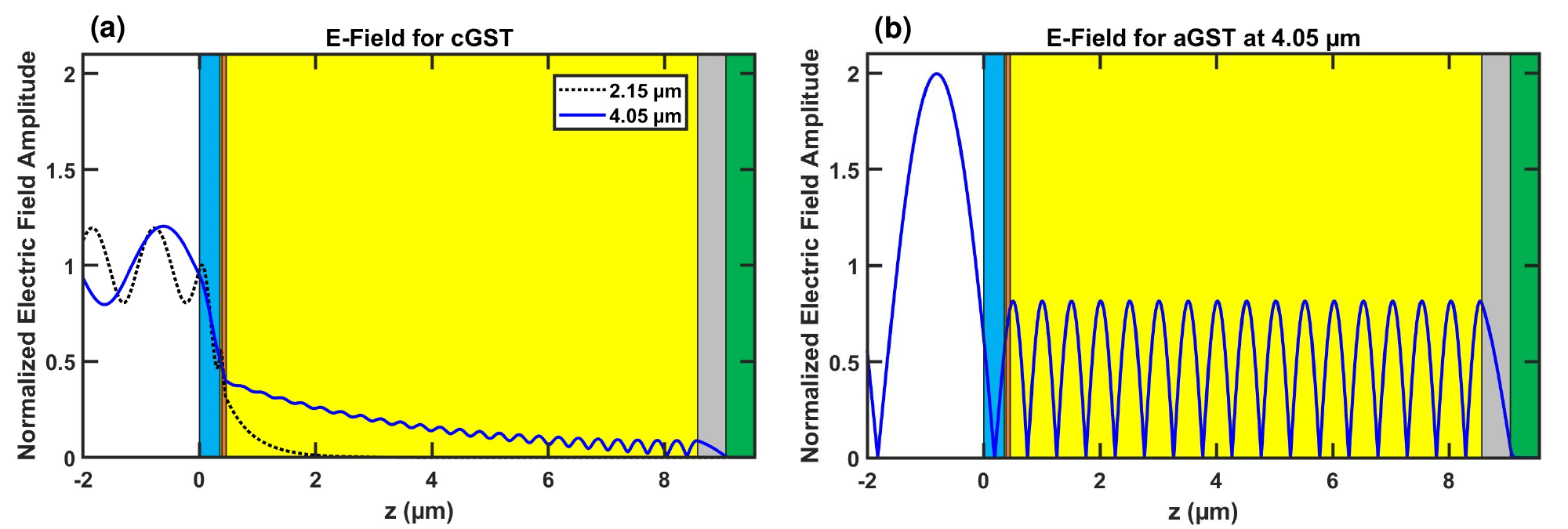} 
\caption{
Profile of the electric field amplitude, normalized with respect to the field amplitude of the normally incident plane wave for the optimized five-layer structure of Table 2. Results are shown for (a) $\lambda_1$ = 2.15 $\mu$m (black dashed line), $\lambda_2$ = 4.05 $\mu$m (blue solid line) with GST in its crystalline phase, and (b) $\lambda_2$ = 4.05 $\mu$m (blue solid line) with GST in its amorphous phase. For GST in its crystalline phase and $\lambda_1$ = 2.15 $\mu$m, the ratio of the power absorbed in each layer to the total absorbed power, beginning with the layer adjacent to air, is \{0, 0.32, 0, 0.68, 0, 0\}. In other words, $\sim$32\% of the total power absorbed is absorbed in the first GST layer, while $\sim$68\% is absorbed in the second GST layer. 
For GST in its crystalline phase and $\lambda_1$ = 4.05 $\mu$m, the ratio of the power absorbed in each layer to the total absorbed power, beginning with the layer adjacent to air, is \{0, 0.03, 0, 0.97, 0, 0\}. In other words, $\sim$3\% of the total power absorbed is absorbed in the first GST layer, while $\sim$97\% is absorbed in the second GST layer.
}
\label{fig:seven}
\end{figure}

We further explain the origin of broadband switchability in the optimized structure by examining the electric field profiles for GST in its crystalline and amorphous phases.
In Figs. \ref{fig:seven}(a) and \ref{fig:seven}(b), we show the profile of the electric field amplitude, normalized with respect to the field amplitude of the normally incident plane wave, for the optimized five-layer structure of Table \ref{table:two}, with GST 
in its crystalline and amorphous phases, respectively.
We observe that, for GST in its crystalline phase, there is almost perfect impedance matching between the structure and air at both $\lambda_1$ = 2.15 $\mu$m and $\lambda_2$ = 4.05 $\mu$m, resulting in very low reflection of $\sim$3.8\% and $\sim$4.2\%, respectively [Fig. ~\ref{fig:seven}(a)]. 
We found that at $\lambda_2$ = 4.05 $\mu$m $\sim$97\% of the total absorbed power is absorbed in the thick 8.104 $\mu$m GST layer, while only $\sim$3\% is absorbed in the thin 0.035 $\mu$m GST layer. However, at $\lambda_1$ = 2.15 $\mu$m $\sim$68\% of the total power absorbed is absorbed in the thick 8.104 $\mu$m GST layer, while $\sim$32\% is absorbed in the thin 0.035 $\mu$m GST layer.
Thus, the thin GST layer
enables our optimized five-layer structure of Table \ref{table:two} to achieve large absorption for GST in its crystalline phase, and therefore large absorption difference between the two GST phases, in a broader wavelength range.
When GST is switched to its amorphous phase, the profile of the electric field amplitude changes drastically, and the reflection increases to $\sim$99.4\% at $\lambda_2$ = 4.05 $\mu$m [Fig. \ref{fig:seven}(b)], and to $\sim$99.2\% at $\lambda_1$ = 2.15 $\mu$m (not shown).

\begin{figure}[htb!]
\centering
\centering\includegraphics[width=11cm]{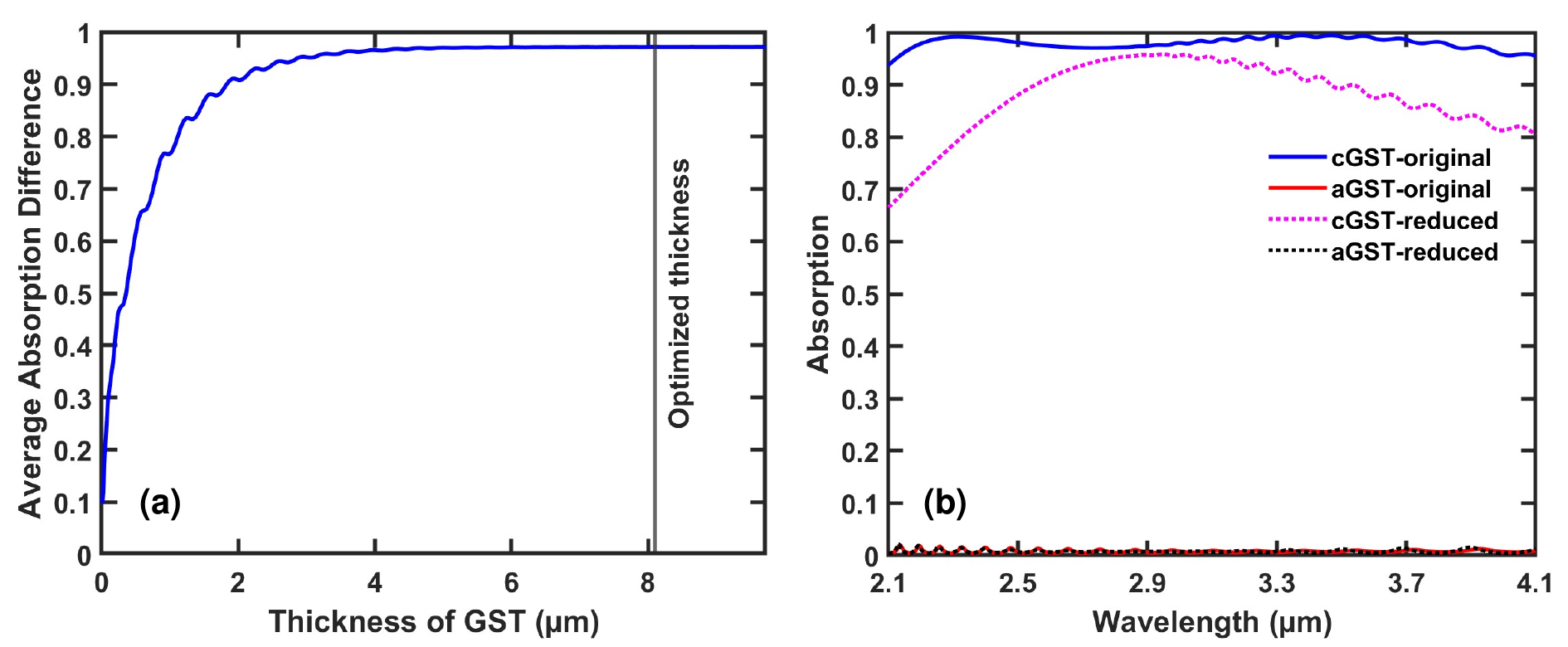} 
\caption{(a) Average difference in the 2.1-4.1 $\mu$m wavelength range between the absorption for GST in its crystalline phase and the absorption for GST in its amorphous phase as a function of the thickness of the second GST layer. All other parameters are as in Table \ref{table:two}. Results are shown for normally incident light.
(b) Absorption as a function of wavelength for the optimized five-layer structure of Table \ref{table:two} with GST in its crystalline (cGST-original) and amorphous (aGST-original) phases. We also show the absorption for a reduced HfO$_2$-GST-MgF$_2$ structure with GST in its crystalline (cGST-reduced) and amorphous (aGST-reduced) phases. The reduced structure is obtained by removing the thin 0.035 $\mu$m GST layer and the SiC layer from the optimized structure of Table \ref{table:two}, while the thicknesses of all remaining layers are the same as in Table \ref{table:two}. Results are shown for normally incident light.
}
\label{fig:eight}
\end{figure}

Figure \ref{fig:eight}(a) shows the average difference in the 2.1-4.1 $\mu$m wavelength range between the absorption for GST in its crystalline phase and the absorption for GST in its amorphous phase as a function of the thickness of the thick GST layer (fourth layer in Table \ref{table:two}).
We observe that the average absorption difference is almost constant for GST layer thickness larger than 4 $\mu$m. More specifically, we find that the average absorption difference is $\sim$96.6\% for a GST layer thickness of 4 $\mu$m, and only slightly increases to $\sim$97.2\% for the optimized GST layer thickness of 8.104 $\mu$m.
In other words, even though the structure obtained from the optimization algorithm has a GST layer with thickness of 8.104 $\mu$m (Table \ref{table:two}), decreasing this thickness by a factor of $\sim$2 to 4 $\mu$m leads to an almost negligible $\sim$0.6\% decrease in the average absorption difference. Thus, the total thickness of the structure in Table \ref{table:two} can be greatly decreased without any significant effect on its performance as a switchable absorber. We also found that decreasing the thick GST layer thickness to 4 $\mu$m does not affect the dependence of the absorption difference on the angle of incidence.

The optimized five-layer structure of Table \ref{table:two} includes a thin and a thick GST layer with thicknesses of 0.035 $\mu$m and 8.104 $\mu$m, respectively. To further elucidate the role of the thin GST layer in the optimized structure, in Fig. \ref{fig:eight}(b) we show the absorption as a function of wavelength with GST in its crystalline (cGST-reduced) and amorphous (aGST-reduced) phases for a reduced three-layer HfO$_2$-GST-MgF$_2$ structure. The reduced structure is obtained by removing the thin 0.035 $\mu$m GST layer and the SiC layer from the optimized structure of Table \ref{table:two}. For comparison, we also show the absorption of the optimized five-layer structure with GST in its crystalline (cGST-original) and amorphous (aGST-original) phases. For GST in its amorphous phase (aGST), the absorption of the reduced three-layer structure is very similar to the absorption of the optimized five-layer structure [Fig. \ref{fig:eight}(b)]. However, for GST in its crystalline phase (cGST) the absorption of the reduced three-layer structure is smaller than the absorption of the optimized five-layer structure in the 2.1-4.1 $\mu$m wavelength range of interest [Fig. \ref{fig:eight}(b)]. This results in a smaller absorption difference for the reduced structure. The decrease in absorption for cGST for the reduced three-layer structure is more pronounced at short wavelengths with the absorption at 2.1 $\mu$m falling below 70\% [Fig. \ref{fig:eight}(b)]. Thus, we conclude that the thin 0.035 $\mu$m GST layer in the optimized structure of Table \ref{table:two} leads to larger difference between the crystalline and the amorphous phase absorption in the 2.1-4.1 $\mu$m wavelength range of interest. This is consistent with the fact that, as mentioned above, at short wavelengths the thin 0.035 $\mu$m GST layer absorbs more than 30\% of the incident power for GST in its crystalline phase [Fig. \ref{fig:seven}(a)].

\begin{figure}[htb!]
\centering
\centering\includegraphics[width=11cm]{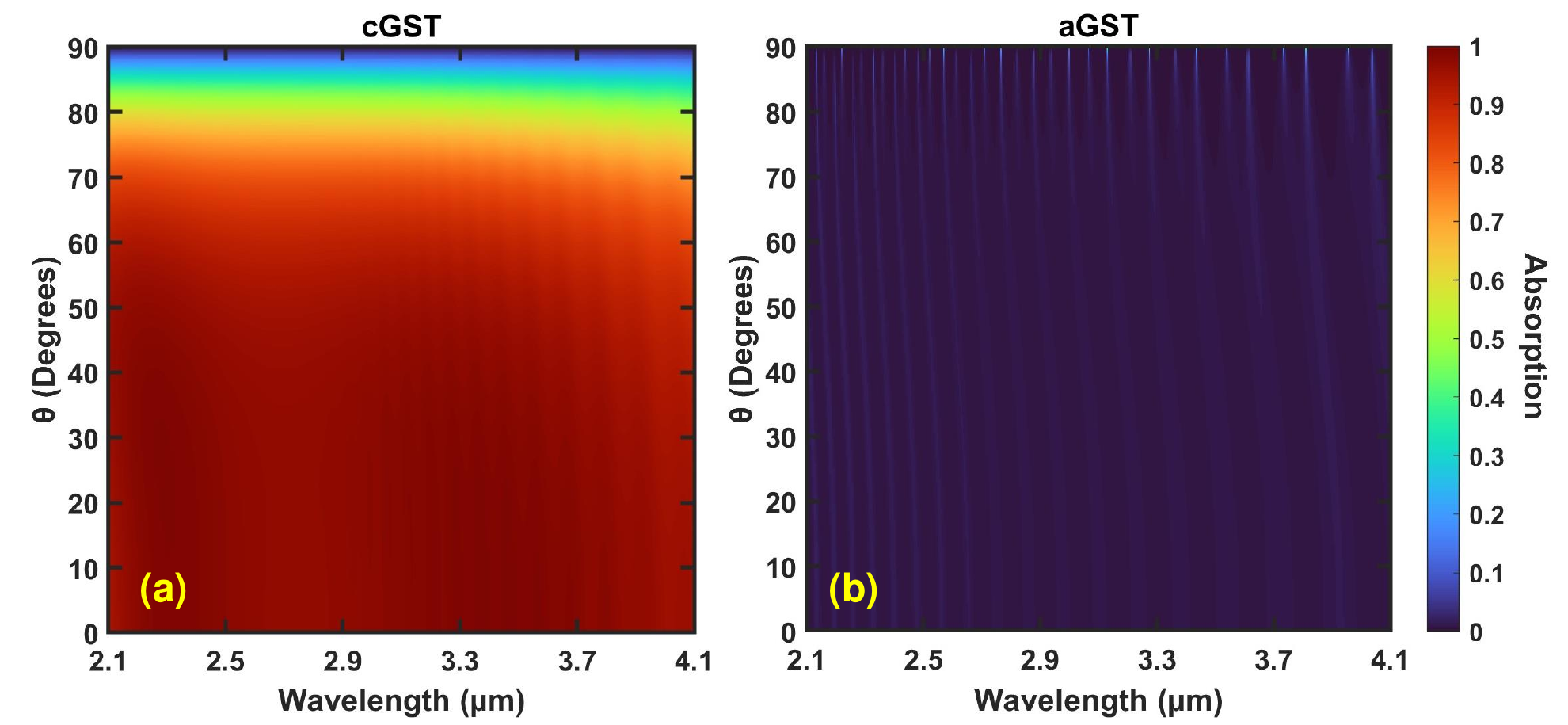} 
\caption{
Absorption as a function of wavelength and angle for the optimized five-layer structure of Table \ref{table:two} with GST in its (a) crystalline and (b) amorphous phases.
}
\label{fig:nine}
\end{figure}

As mentioned above, the structure of Table \ref{table:two} was obtained by maximizing the absorption difference between the two GST phases in the 2.1-4.1 $\mu$m wavelength range for normally incident light. We now also investigate the dependence of the absorption difference on the angle of incidence. In Figs. \ref{fig:nine}(a) and \ref{fig:nine}(b) we show the absorption as a function of wavelength and angle of incidence for the optimized five-layer structure of Table 2 with GST in its crystalline and amorphous phases, respectively. As discussed in Subsection \ref{1to1.6}, we expect that the absorption spectra will not have a strong angular dependence. We indeed observe that the difference between the absorption for GST in its crystalline phase [Fig. \ref{fig:nine}(a)] and the absorption for GST in its amorphous phase [Fig. \ref{fig:nine}(b)] is high in a broad angular range within the 2.1-4.1 $\mu$m wavelength range of interest. 
More specifically, the absorption for GST in its crystalline phase exceeds 90\%
for incident angles up to $70^\circ$ [Fig. \ref{fig:nine}(a)]. In addition, the absorption for GST in its amorphous phase remains close to zero at all angles of incidence [Fig. \ref{fig:nine}(b)]. Thus, the optimized five-layer structure of Table \ref{table:two} achieves large absorption difference between the two GST phases both in a broad wavelength and in a broad angular range.
We also note that the angular range of large absorption difference is broader for the structure of Table \ref{table:two} optimized in the 2.1-4.1 $\mu$m wavelength range (Fig. \ref{fig:nine}), compared to the structure of Table \ref{table:one} optimized in the 1-1.6 $\mu$m wavelength range (Fig. \ref{fig:five}). As mentioned above, the superior performance of the 2.1-4.1 $\mu$m switchable absorber is directly related to the optical properties of GST. 

\section{Conclusions}
\label{Conclusion}

In this paper, we introduced multilayer structures with the phase-change material GST above a semi-infinite metal substrate for use as broadband switchable absorbers in the infrared wavelength range. We designed switchable absorbers operating in two broad wavelength ranges in the infrared: 1-1.6 $\mu$m and 2.1-4.1 $\mu$m. Our goal is to maximize the difference between the absorption for GST in its crystalline phase and the absorption for GST in its amorphous phase in the wavelength range of interest. In other words, the ideal structure has complete absorption for GST in its crystalline phase, and complete reflection for GST in its amorphous phase. 

We used a memetic optimization algorithm coupled with the impedance method to optimize both the material composition and the layer thicknesses of the multilayer structures, in order to minimize the merit function of Eq. (2). The material of each of the five layers was chosen among Al$_2$O$_3$, HfO$_2$, MgF$_2$, SiC, Si$_3$N$_4$, SiO$_2$, Ta$_2$O$_5$, TiO$_2$, and GST. The substrate was chosen as silver and the superstrate as air.

We first employed the memetic optimization algorithm to design a switchable absorber operating in the 1-1.6 $\mu$m wavelength range. Our optimized structure consists of a thin GST film on top of a MgF$_2$ dielectric spacer backed with an Ag mirror. The structure also includes three additional dielectric layers (MgF$_2$-SiC- Si$_3$N$_4$) on top of the GST layer.
Despite the small thickness of the GST layer in the optimized structure, switching between the crystalline and amorphous phases of GST leads to drastic changes in the electric field profile.
In addition, the average absorption difference is highly dependent on the thickness of the thin GST layer. 
The three dielectric layers on top of the GST layer enable our optimized five-layer structure to achieve large absorption difference between the two GST phases in a broader wavelength and angular range.
We found that the average absorption in the 1-1.6 $\mu$m wavelength range for GST in its crystalline phase is $\sim$93.6\%, while for GST in its amorphous phase it is reduced to $\sim$25.8\%.
Although the multilayer structure was optimized for normally incident light, it achieves large absorption difference between the two GST phases in a broad angular range.
Our optimized lithography-free five-layer structure has better performance than three-dimensional structures reported in the literature which require nanoimprint or electron beam lithography.

We also employed the memetic optimization algorithm to design a switchable absorber operating in the 2.1-4.1 $\mu$m wavelength range. Our optimized structure includes a thin and a thick GST layer, separated by a SiC layer. In addition, a MgF$_2$ dielectric layer separates the bottom GST layer from the silver substrate, while an HfO$_2$ layer separates the top GST layer from air.
At short wavelengths the thin GST layer absorbs more than 30\% of the incident power for GST in its crystalline phase. The thin GST layer therefore enables our optimized five-layer structure to achieve large absorption for GST in its crystalline phase, and therefore large absorption difference between the two GST phases, in a broader wavelength range.
We found that the average absorption in the 2.1-4.1 $\mu$m wavelength range for GST in its crystalline phase is $\sim$98.0\%, while for GST in its amorphous phase it is reduced to $\sim$0.8\%. Thus, our optimized structure approaches the ideal structure, since it is almost a perfect absorber for GST in its crystalline phase and almost a perfect reflector for GST in its amorphous phase in the 2.1-4.1 $\mu$m wavelength range of interest.
In addition, the angular range of large absorption difference is broader for the 2.1-4.1 $\mu$m switchable absorber compared to the 1-1.6 $\mu$m switchable absorber.
The superior performance of the 2.1-4.1 $\mu$m switchable absorber compared to the 1-1.6 $\mu$m switchable absorber is directly related to the optical properties of GST. Amorphous GST is almost lossless for wavelengths longer than 2 $\mu$m, while crystalline GST is lossy in the same wavelength range. This enables the design of a close to ideal broadband switchable absorber for wavelengths longer than 2 $\mu$m, with almost perfect absorption for the lossy crystalline GST, and almost perfect reflection for the lossless amorphous GST. 

As final remarks, our optimized switchable absorbers have five layers and are based on the phase-change material GST (Tables 1 and 2). 
We also optimized structures with more than five layers, but found that increasing the number of layers beyond five does not lead to a significant increase in the average absorption difference between the two GST phases in the wavelength ranges of interest. In addition, even though we also optimized multilayer structures based on the phase-change material VO$_2$, we found that GST-based structures have better performance than VO$_2$-based structures in terms of the average absorption difference in the wavelength ranges of interest.

Our results demonstrate the effectiveness of the proposed multilayer structures in achieving switching between near perfect absorption and very low absorption in a broad wavelength range. 
They could pave the way to a new class of broadband switchable absorbers and thermal sources in the infrared wavelength range.
Our proposed structures have potential applications in thin-film thermal emitters, thermo-solar cells, photodetectors, and smart windows. 

\section*{Funding}
National Science Foundation (1254934, 2150491).

\section*{Disclosures}
The authors declare no conflicts of interest.

\section*{Data Availability}
Data underlying the results presented in this paper are not publicly available at this time but may be obtained from the authors upon reasonable request.

\bibliography{reference}

\end{document}